\documentclass[cits]{PoS}
\title{Axion helioscopes update: the status of CAST \& IAXO}
\ShortTitle{Axion helioscopes update: the status of CAST \& IAXO}
\author{T.~Dafni and \speaker{F.J.~Iguaz}, on behalf of the CAST and IAXO collaborations\\
        Laboratorio de F\'isica Nuclear y Astropart\'iculas, Universidad de Zaragoza, Spain.\\
        E-mail: \email{tdafni@cern.ch} and \email{iguaz@cern.ch}}

\abstract{Almost 35 years since their suggestion as a good solution to the strong CP-problem, axions
remain one of the few viable candidates for the Dark Matter, although still eluding detection. Most of
the methods for their detection are based on their coupling to photons, one of the most sensitive
ones being the helioscope technique. We report on the current status of the CERN Axion
Solar Telescope and the future International Axion Observatory (IAXO). Recent results from the
second part of CAST phase II, where the magnet bores were filled with $^3$He gas at variable pressure
achieving sensibilities on the axion mass up to 1.2 eV, are presented. Currently, CAST is expecting
to improve its sensitivity to solar axions with rest mass below 0.02 eV/c$^2$ after the upgrade of the
X-ray detectors and with the implementation of a second X-ray optic. At the same time, it is
exploring other possibilities at the low energy physics frontier. On the other hand IAXO, the fourth
generation axion helioscope, aims to improve CAST's performance in terms of axion-photon coupling
by 1-1.5 orders of magnitude. The details of the project building a dedicated magnet, optics and
X-ray detectors are given.}

\FullConference{Technology and Instrumentation in Particle Physics 2014,\\
		2-6 June, 2014\\
		Amsterdam, the Netherlands}

\begin{document}
\section{Introduction}
Axions are hypothetical particles that have been postulated in models that may explain the strong-CP problem \cite{Wantz:2010}.
They are practically stable, and may have been abundantly produced in the early universe,
which makes them a good candidate for the Dark Matter \cite{Archidiacono:2013cha}.
The unknown symmetry scale $f_a$ determines the axion phenomenology
and the mass can be expressed in the form m$_a = 6~{\rm eV}(10^{6}~{\rm GeV}/f_a)$. 
Most experimental searches for the axion are based on their interaction with two photons,
allowed in all models, characterized by the coupling constant $g_{a\gamma}$.
As a consequence, axions can transform into photons and vice-versa in strong external electric and magnetic fields,
the Primakoff process. This process points to the stars producing axions through the transformation of thermal photons
in the presence of the electric fields of the nuclei and the stellar plasma.
The Sun being our closest star would be the brightest source of axions
and is the reason for the helioscope technique \cite{Sikivie:1983ip}:
solar axions streaming from the Sun would be reconverted into photons
in the presence of a transverse electromagnetic field in a laboratory.

\medskip
The expected signal is in the energy range of 1--10~keV. The excess of photons in this energy range when looking
at the Sun, i.e. in axion sensitive conditions over the background X-rays
during the rest of the time would constitute a positive signal.
The number of excess photons expected depends on the very weak $g_{a\gamma}$ directly related to the sensitivity
~\cite{Irastorza:2011gs} as seen in the following equation
\begin{equation}
g_{a\gamma}^4 \sim B^2 L^2 A \; \epsilon_d b^{-1/2} \; \epsilon_o a^{-1/2} \; \epsilon_t^{1/2} t^{1/2}
\label{sens}
\end{equation}
\noindent From \ref{sens} become apparent the factors that drive the sensitivity of a helioscope:
the length $L$ and the strength $B$ of the provided magnetic field and the axion-sensitive area $A$
are factors related to the magnet;
the background level $b$ and the efficiencies $\epsilon_d$ of the X-ray detectors used;
the efficiency $\epsilon_o$ and total focusing area $a$ of focusing optics,
the use of which significantly enhances the signal-to-background ratio of the helioscope
and finally, the fraction of time the magnet tracks the Sun $\epsilon_t$
and the total time of data-taking of the experiment $t$.

\medskip
The first implementation of this technique was the Rochester-Brookhaven-Florida experiment \cite{rbf1,rbf2}.
Later a more sensitive search was performed by SUMICO \cite{Moriyama:1998kd,Inoue:2002qy,Inoue:2008zp}.
The most sensitive helioscope to date is CAST.
The IAXO project plans to improve the sensitivity of CAST in the near future.
We shortly review here CAST's latest results \cite{Arik:2014}
and summarise the IAXO papers \cite{Irastorza:2011gs, loi, Irastorza:2014} to which the reader is referred for further details.

\section{CAST}
The CERN Axion Solar Telescope (CAST) was the first axion helioscope to implement an X-ray focusing device
in front of a CCD camera, improving the experiment's sensitivity. The other major improvement by CAST was
the use of a decommissioned LHC prototype magnet, which provided a 9~T field over approximately 10~m.
The magnet has a twin aperture and allowed for the use of four detectors,
compensating in this way the small aperture of the accelerator magnet ($\sim 14$ cm$^2$),
which can follow the Sun during 1.5 h at sunrise and at sunset,
thanks to the moving platform on which the magnet is sitting.
The other three bores have been occupied by a Micromegas detector and a conventional Time Projection Chamber
which was later replaced by two more Micromegas detectors.

\begin{figure}[htb!]
 \begin{center}
\includegraphics[width=0.7\columnwidth]{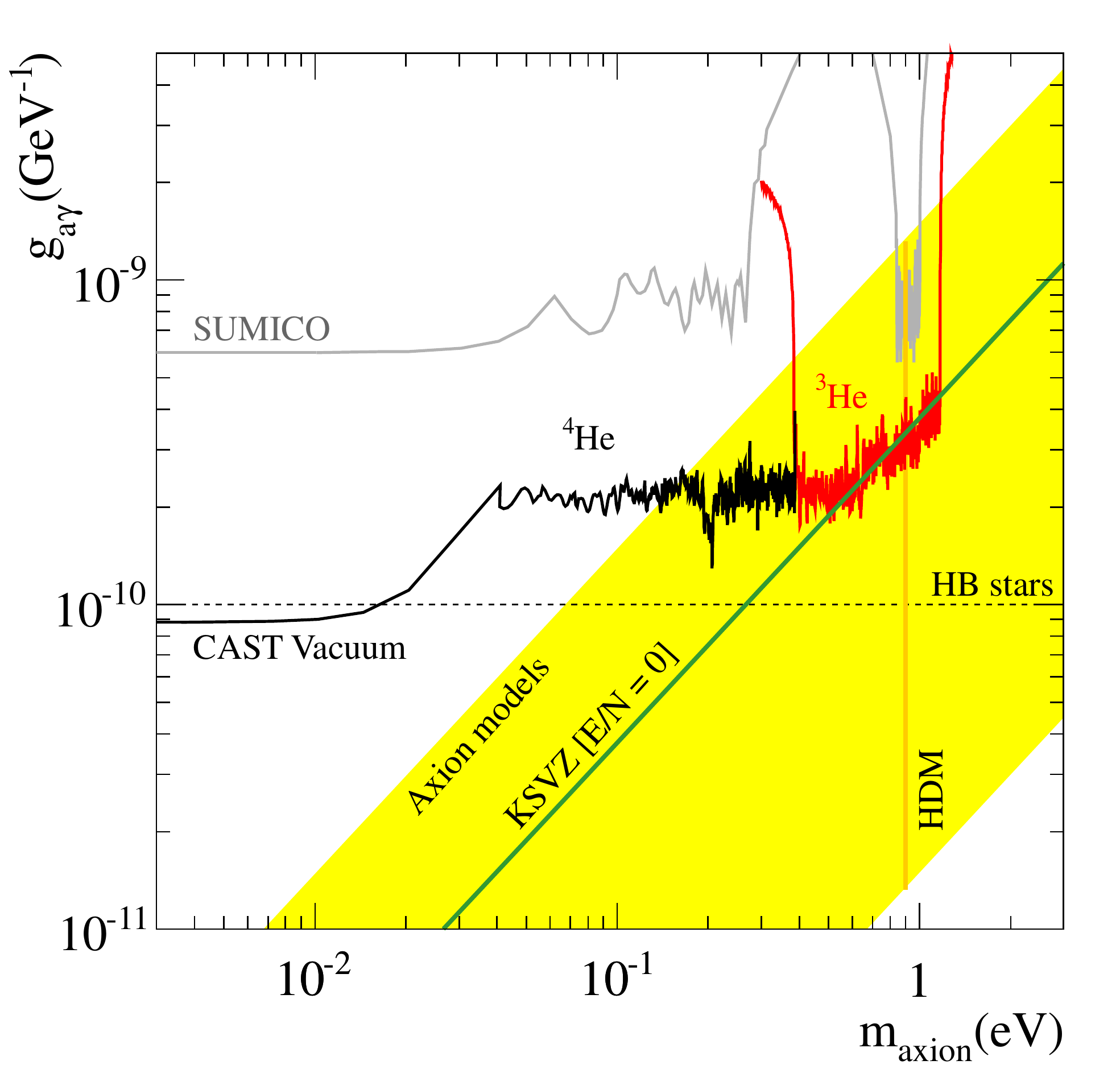}
\end{center}
\caption{Expanded view of the CAST results in the vacuum,$^4$He (in black) and the $^3$He phases
with the new limit (in red). The limit from SUMICO \cite{Inoue:2008zp},
the hot dark matter (HDM) bound \cite{Archidiacono:2013cha}
and the horizontal brunch (HB) stars \cite{Raffelt:2008ggr} are also shown. The yellow band denotes typical theoretical models,
while the green solid line corresponds to $E/N=0$ (KSVZ model).}
\label{fig:cast}
\end{figure}

\medskip
Between 2003 and 2013, CAST gathered data in different experimental conditions,
putting the most stringent limits on the axion coupling constant at the masses which it covered,
as can be seen in Fig. \ref{fig:cast}. Axion masses up to m$_{\rm a}\sim$0.02~eV were covered
during the first phase of CAST, with the magnet bores operated under vacuum ~\cite{Zioutas:2004hi, Andriamonje:2007ew}.
Later on, the system was modified in order to fill the bores with a gas;
the presence of the gas inside the magnet restores the coherence of the axion and photon waves
and therefore maintains the conversion probability at high levels for axion masses larger than 0.02 eV.
Increasing the density of $^4$He inside the magnet in steps, CAST scanned axion masses 
up to $m_{\rm a}\sim$0.39~eV~\cite{Arik:2008mq}, entering for the first time in the QCD-favoured band.
In order to push the sensitivity to even higher masses, and given that $^4$He condensates at higher pressures,
CAST started injecting $^3$He inside the bores, covering as a first step masses
up to m$_{\rm a}\sim$0.64~eV~\cite{Arik:2011rx} and finally reaching $m_{\rm a}\sim$1.17~eV.
No evidence of a signal was found, leading to an upper bound of the axion-to-photon constant
of $g_{a\gamma}<3.3 \times 10^{-10}$ GeV$^{-1}$ for the mass range between 0.64~eV and 1.17~eV~\cite{Arik:2014}.

\medskip
In parallel to its main physics program, CAST has performed other searches related to axions
including high energy axions with a $\gamma$-ray calorimeter \cite{Andriamonje:2009ar},
low-energy axions in the visible \cite{Cantatore:2008ue}
and 14.1~keV axions from M1 nuclear transitions at the Sun \cite{Andriamonje:2009dx}.
Studying the axion-to-electron coupling, allowed in non-hadronic axion models,
CAST has also published constraints on the combined axion-electron and axion-photon coupling constants~\cite{Barth:2013sma}.

\medskip
After concluding the $^3$He phase, CAST decided to go back to the $^4$He (in 2012) and the vacuum phase (at present);
in the latter CAST has been able to also look at the low energy part for evidence of other particles such as chameleons,
which appear in Dark Energy models, or hidden photons \cite{Jaeckel:2010ni}.
The decision to revisit these regions was made because the sensitivity of the experiment was improved over the last years.
This improvement is due to the low-background Micromegas detectors which have reached levels
down to $7 \times 10^{-7}$~keV$^{-1}$~cm$^2$~s$^{-1}$ \cite{Garza:2014jg} after dedicated efforts regarding the shielding
and the low radioactivity materials in the vicinity of the detectors \cite{Aune:2014sa}.
In 2014, one of the Micromegas detectors will be equipped with a new X-ray-focusing device,
which is expected to increase the signal-to-noise ratio significantly and
is a pathfinder project in view of the International AXion Observatory (IAXO). 

\section{IAXO}
The IAXO project proposes a fourth generation helioscope which will further improve the sensitivity of the previous ones,
probably up to the limit this technique can offer.
Helioscopes so far have used recycled material (magnets, telescopes)
that was not optimized for axion searches and IAXO will address exactly this shortcoming,
improving on all aspects related to the sensitivity of the helioscopes. In combination with ADMX~\cite{admx2},
IAXO can explore a big part of the axion model in the next decade and, as later described, there is a big
potential for new physics.

\medskip
A Letter of Intent has already been presented to CERN \cite{loi} along
with a Conceptual Design Report \cite{Irastorza:2014} of the experiment. The letter has received
a positive recognition of the physics motivation and a confirmation that the proposed design is appropriate
to the use of state-of-art technologies.
The IAXO community is now developing several preparatory activities towards a Technical Design Report,
which consist in different prototypes of the future magnet coils, optics and detectors.

\begin{figure}[htb!]
\centering
\includegraphics[width=0.49\columnwidth]{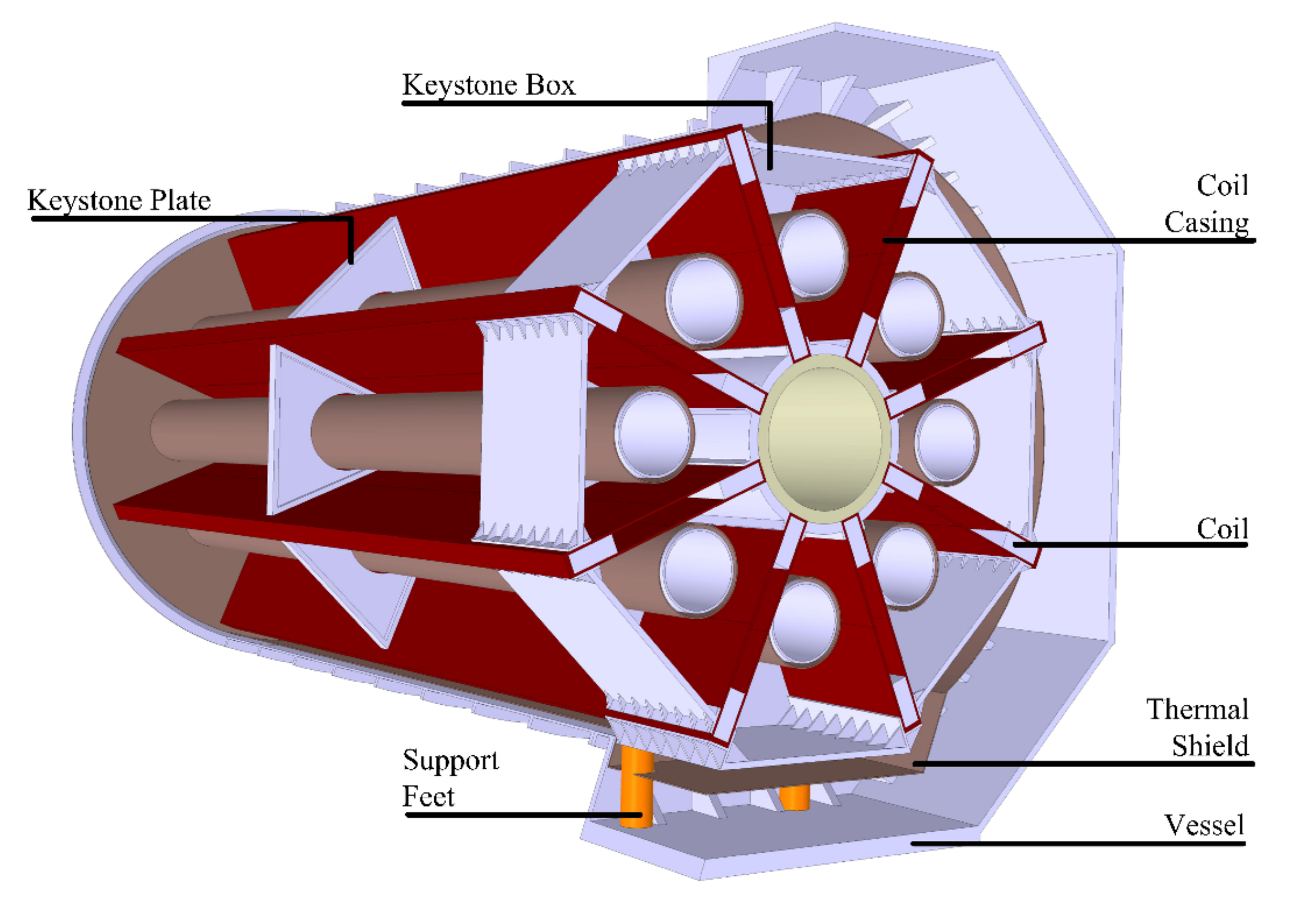}
\includegraphics[width=0.49\columnwidth]{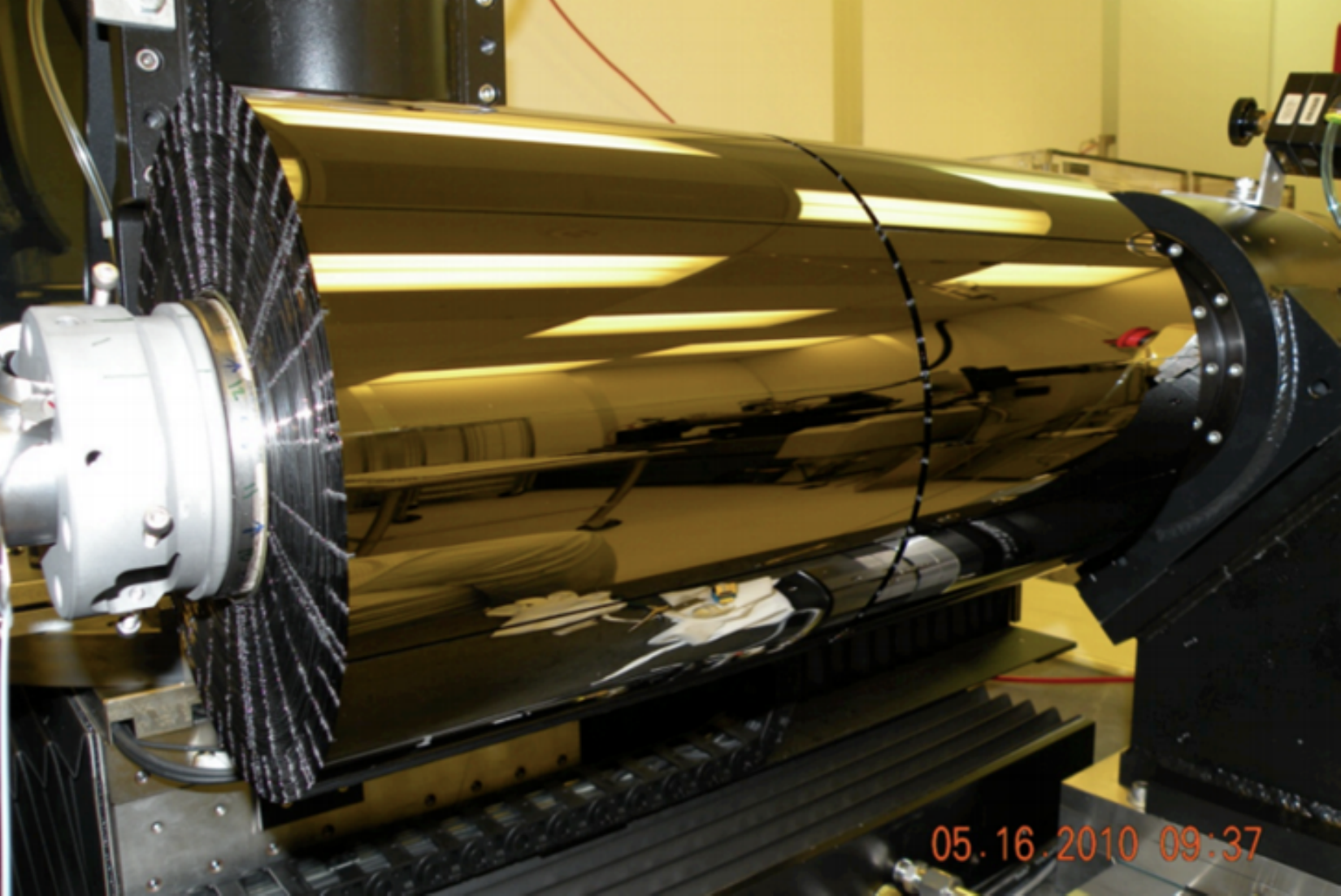}
\caption{Left: Mid-plane cut of the cryostat showing the cold mass and its supports,
surrounded by a thermal shield, and the vacuum vessel \cite{Shilon:2012}. Right: A view of the the focusing device developed
for the NuSTAR satellite mission \cite{Madsen:2009kkm}.}
\label{fig:iaxoparts}
\end{figure}

\medskip
The IAXO collaboration proposes the development of a dedicated magnet \cite{Shilon:2012}, built with existing technology,
which will optimize the features of the magnet as seen in eq. \ref{sens}.
The new magnet is conceived as an ATLAS-like toroidal (as shown in Fig. \ref{fig:iaxoparts}, left)
which will be able to deliver up to 2.5~T along its 25~m length;
the main improvement will come from the bore's aperture of this design,
as 8 bores with a diameter close to 60~cm will increase the magnet figure of merit by a factor 300 \cite{Irastorza:2011gs}.
The magnet is planned to be sitting on a movable platform which will be pointing to the Sun for approximately 12~h per day.

\medskip
Eight X-ray focusing devices are foreseen to cover the eight bores of the magnet.
Based on the technology already developed for the NuSTAR satellite mission \cite{Madsen:2009kkm},
again IAXO is counting on the experience of the collaborators on delivering telescopes
(see Fig. \ref{fig:iaxoparts}, right)
that will cover the requirements for IAXO maximizing the throughput (area of 2800 cm$^2$)
and minimizing the focal spot (area of 16 mm$^2$).
The first prototype of these optics coupled to a Micromegas detector will be soon installed at CAST.

\medskip
Micromegas detectors of the microbulk type will be the baseline X-ray detectors of IAXO \cite{Aune:2014sa}.
These detectors show a high power to discriminate X-rays signals from background, are intrinsic radiopure
and shielding techniques from low background experiments can be applied as well.
The performance of these detectors in CAST but also at other dedicated test benches at sea level 
and at the Canfranc Underground Laborator shows that the background requirements of IAXO
of $10^{-7}$~keV$^{-1}$~cm$^2$~s$^{-1}$ (or better) are not far from reach \cite{Garza:2014jg}.
The small size of the spot allows a reduced sensitive area of the detectors and therefore
a small volume which can be efficiently covered by an appropriate shielding.
The pathfinder will be equipped with a complete shielding
and will provide information on various parameters, as well as valuable experience of the operation.

\begin{figure}[htb!]
 \begin{center}
\includegraphics[width=0.7\columnwidth]{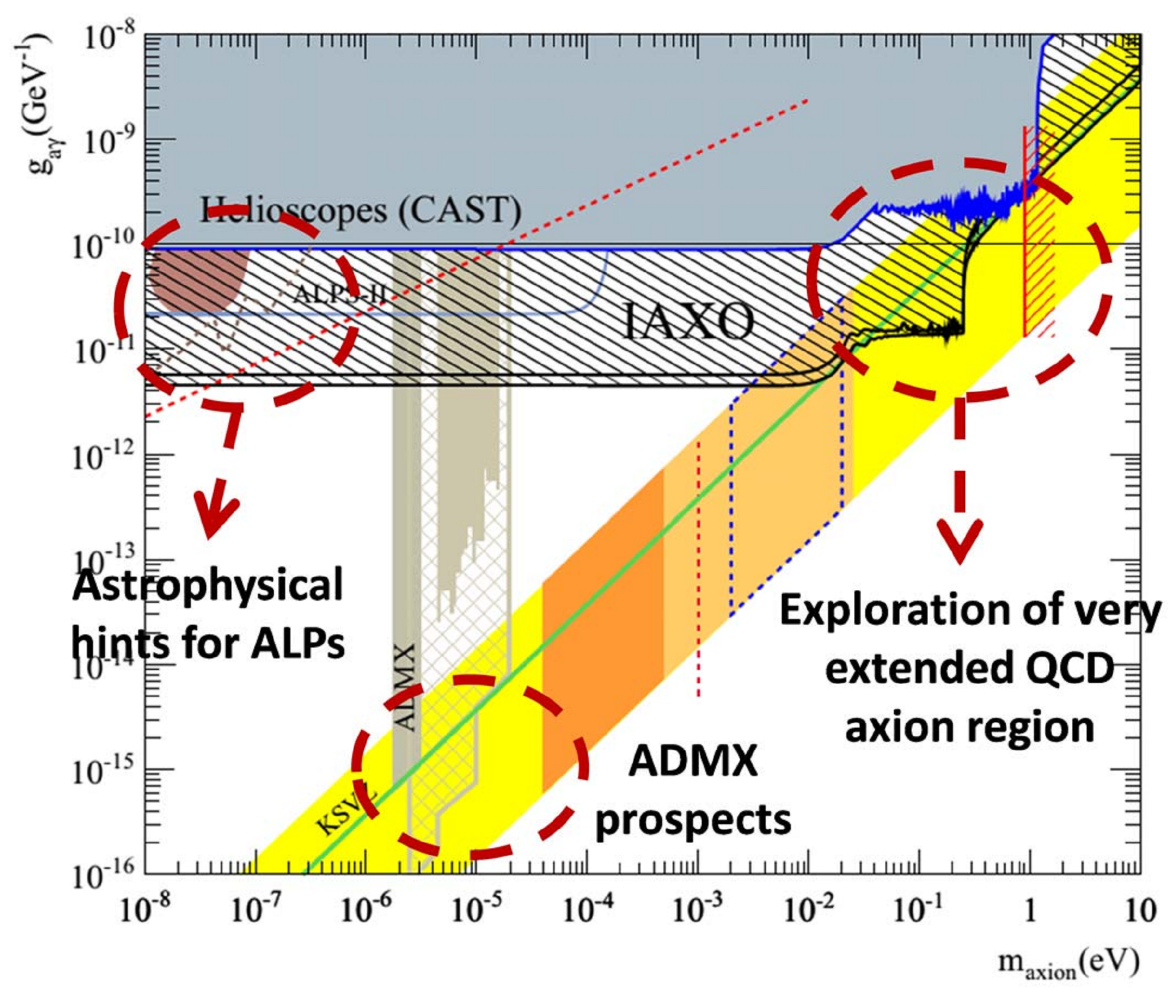}
\end{center}
\caption{An extended version of the axion exclusion plot with the expected IAXO sensitivity,
where it is compared with current bounds (solid)
and future prospects (dashed) of other experiments (CAST, ADMX~\cite{admx2}, ALPS-II~\cite{alpsII}).
The solid brown region excluded by H.E.S.S. data~\cite{hess}, is within the low m$_a$ region invoked
in the context of the transparency of the universe, denoted with the dashed grey line.
Below the red dashed line lies the region in which ALP DM parameter space is viable.
We refer to ~\cite{loi} for a more detailed explanation of the plot.}
\label{fig:iaxo}
\end{figure}

\medskip
The expected improvement achieved after these optimizations on the crucial equipment of IAXO sum
up to a S/B ratio of 5 orders of magnitude better than CAST,
which leads to a sensitivity of $g_{a\gamma}$ down to few times $10^{-12}$ GeV$^{-1}$
for masses up to about 0.02~eV, entering in the QCD axion band as it can be seen in Fig. \ref{fig:iaxo}.
IAXO will push the sensitivities to the limits of the helioscope technique when looking for axions
and ALPs coming from the Sun, its primary physics goal.
Modifying the system in a way to allow the introduction of gas inside the bores would be an additional step
in order to cover higher axion masses, up to the $\sim$1~eV region.
In fact, various models in which the axion appears as part of the hot dark matter of the Universe could be probed.

\medskip
IAXO can also provide a definitive test of the existence of Axion-Like Particles (ALPs) with very low masses,
below 10$^{-7}$~eV, which have been proposed as an explanation to anomalies
in light propagation over astronomical distances~\cite{Meyer:2013pny}.
The study of the axion-to-electron coupling is another possibility.
These studies have sparked more interest since axions with $g_{ae}$ of few $\times 10^{-13}$ could explain
the anomalous cooling from astrophysical observations in white dwarfs~\cite{Isern:2010}
and IAXO would have enough sensitivity to measure the solar axion flux produced with the same mechanism.

\medskip
IAXO can also serve as a platform for studies of other proposed particles such as hidden photons or chameleons,
introduced in the context of Dark Energy. The design of the IAXO magnet would also allow the incorporation
of other equipment inside the bores, such as microwave cavities or antennas,
to initiate searches for the direct detection of relic axions that would have been produced
in the early stages of the Universe, in parallel to the baseline program of the project. 

\section{Conclusions}
The CAST experiment has published its final results from the $^3$He phase,
during which it entered well into the QCD-axion region at the upper masses end.
No excess of photons over the background was observed, whichd lead to an upper limit
for the axion-photon coupling constant of $g_{a\gamma} < 3.3 \times 10^{-10}$ GeV$^{-1}$
for axion masses between 0.64~eV and 1.17~eV.
Currently CAST is revisiting the vacuum phase with increased sensitivity and
is looking for other exotica such as hidden photons and chameleons.
In parallel it is installing an X-ray focusing device coupled to a Micromegas detector
which will serve as an important benchmark for the IAXO project.
IAXO is expecting to improve the sensitivity reached by CAST by more than an order of magnitude
using a dedicated magnet and X-ray focusing devices for all 8 bores of the magnet,
coupled to very low background Micromegas detectors.
It is also exploring the possibility of enlarging its axion-physics program in order to include relic-axion searches.

\section*{Acknowledgements}
We thank CERN for hosting the CAST experiment and for the technical support to operate the magnet and the cryogenics.
We thank the CERN CFD team for their essential contribution to the CFD work.
We acknowledge support from the Spanish Ministry of Science and Innovation (MICINN) under
contract FPA2008-03456 and FPA2011-24058, as well as under the CPAN project CSD2007-00042
from the Consolider-Ingenio2010 program of the MICINN. Part of these grants are funded by the
European Regional Development Fund (ERDF/FEDER). We also acknowledge support from the
European Commission under the European Research Council T-REX Starting Grant ERC-2009-StG-240054
of the IDEAS program of the 7th EU Framework Program. Part of this work was
performed under the auspices of the U.S. Department of Energy by Lawrence Livermore National
Laboratory under Contract DE-AC52-07NA27344 with support from the LDRD program through
grant 10-SI-015. The design work on the IAXO magnet system was supported by CERN, Physics Department as well as
the ATLAS Collaboration. Partial support by the Deutsche Forschungsgemeinschaft (Germany) under grant EXC-153,
by the MSES of Croatia and the Russian Foundation for Basic Research (RFBR) is also acknowledged.
F.I. acknowledges the support from the \emph{Juan de la Cierva} program.


\begin{thebibliography}{99}
\bibitem{Wantz:2010}
O.~Wantz and E.~P.~S.~Shellard, 
\emph{Axion Cosmology Revisited}, 
\emph{Phys.\ Rev.\ D} {\bf 82}, 123508(2010) [arXiv:0910.1066]

\bibitem{Archidiacono:2013cha}
M.~Archidiacono, S.~Hannestad, A.~Mirizzi, G.~Raffelt and Y.~Y.~Y.~Wong,
\emph{Axion hot dark matter bounds after Planck},
[arXiv:1307.0615].

\bibitem{Sikivie:1983ip}
P.~Sikivie, 
\emph{Experimental tests of the ``invisible'' axion},
\emph{PRL} {\bf 51}, 1415 (1983);
(E) {\it ibid.}\ {\bf 52}, 695 (1984).

\bibitem{Irastorza:2011gs} 
I.~G.~Irastorza {\it et al.},
\emph{Towards a new generation axion helioscope},
\emph{JCAP} {\bf 1106}, 013 (2011) [arXiv:1103.5334 [hep-ex]].

\bibitem{rbf1} 
K.~van Bibber {\it et al.}, 
\emph{Design for a practical laboratory detector for solar axions}, 
\emph{Phys.\ Rev.\ D} {\bf 39}, 2089 (1989).

\bibitem{rbf2} 
D.~M.~Lazarus et al.,
\emph{Search for solar axions}, 
\emph{PRL} {\bf 69}, 2333 (1992).

\bibitem{Moriyama:1998kd}
S.~Moriyama {\it et al.},
\emph{Direct search for solar axions by using strong magnetic field and X-ray  detectors},
\emph{Phys.\ Lett.\  B} {\bf 434}, 147 (1998).

\bibitem{Inoue:2002qy}
Y.~Inoue {\it et al.},
\emph{Search for sub-electronvolt solar axions using coherent conversion of axions
into photons in magnetic field and gas helium},
\emph{Phys.\ Lett.\  B} {\bf 536}, 18 (2002)[arXiv:astro-ph/0204388].

\bibitem{Inoue:2008zp}
Y.~Inoue {\it et al.},
\emph{Search for solar axions with mass around 1 eV using coherent conversion of axions into photons},
\emph{Phys.\ Lett.\  B} {\bf 668}, 93 (2008)[arXiv:0806.2230 [astro-ph]].

\bibitem{Arik:2014} 
M.~Arik {\it et al.}  [CAST Collaboration],
\emph{Search for Solar Axions by the CERN Axion Solar Telescope with $^{3}\mathrm{He}$ Buffer Gas:
Closing the Hot Dark Matter Gap},
\emph{PRL} {\bf 112}, 091302 (2014) [arXiv:1307.1985 [hep-ex]].  

\bibitem{loi}
I.G.~Irastorza {\it et al.},
\emph{The International Axion Observatory IAXO, Letter of Intent to the CERN SPS commitee},
CERN-SPSC-2013-022, SPSC-I-242 (2013).

\bibitem{Irastorza:2014} 
E.~Armengaud {\it et al.},
\emph{Conceptual design of the International Axion Observatory (IAXO)},
\emph{JINST} {\bf 9}, T05002 (2014).

\bibitem{Zioutas:2004hi} 
K.~Zioutas {\it et al.}  [CAST Collaboration],
\emph{First results from the CERN Axion Solar Telescope (CAST)}, 
 \emph{PRL} {\bf 94}, 121301 (2005)  [hep-ex/0411033]. 

\bibitem{Andriamonje:2007ew} 
S.~Andriamonje {\it et al.}  [CAST Collaboration],
\emph{An Improved limit on the axion-photon coupling from the CAST experiment},  
\emph{JCAP} {\bf 0704}, 010 (2007) [hep-ex/0702006].

\bibitem{Arik:2008mq} 
E.~Arik {\it et al.}  [CAST Collaboration],
\emph{Probing eV-scale axions with CAST},  
\emph{JCAP} {\bf 0902}, 008 (2009) [arXiv:0810.4482 [hep-ex]].

\bibitem{Arik:2011rx} 
M.~Arik {\it et al.}  [CAST Collaboration],
\emph{CAST search for sub-eV mass solar axions with $^3$He buffer gas},  
\emph{PRL} {\bf 107}, 261302 (2011) [arXiv:1106.3919 [hep-ex]].

\bibitem{Andriamonje:2009ar} 
S.~Andriamonje {\it et al.}  [CAST Collaboration],
\emph{Search for solar axion emission from $^7Li$ and $D(p,\gamma)^3He$ nuclear decays with the CAST $\gamma$-ray calorimeter},
\emph{JCAP} {\bf 1003}, 032 (2010) [arXiv:0904.2103 [hep-ex]].

\bibitem{Cantatore:2008ue} 
G.~Cantatore {\it et al.}  [CAST Collaboration],
\emph{Search for low Energy solar Axions with CAST}, [arXiv:0809.4581 [hep-ex]].

\bibitem{Andriamonje:2009dx} 
S.~Andriamonje {\it et al.}  [CAST Collaboration],
\emph{Search for 14.4-keV solar axions emitted in the M1-transition of Fe-57 nuclei with CAST},
\emph{JCAP} {\bf 0912}, 002 (2009) [arXiv:0906.4488 [hep-ex]].
  
\bibitem{Barth:2013sma} 
K.~Barth {\it et al.},
\emph{CAST constraints on the axion-electron coupling},
\emph{JCAP} {\bf 1305}, 010 (2013) [arXiv:1302.6283 [astro-ph.SR]].

\bibitem{Jaeckel:2010ni} 
J.~Jaeckel and A.~Ringwald,
\emph{The Low-Energy Frontier of Particle Physics},  
\emph{Ann.\ Rev.\ Nucl.\ Part.\ Sci.\ } {\bf 60}, 405 (2010)  [arXiv:1002.0329 [hep-ph]].

\bibitem{Garza:2014jg}
J.G.~Garza {\it et al.},
\emph{X-ray detection with Micromegas with background levels below 10$^{-6}$ keV$^{-1}$ cm$^{-2}$ s$^{-1}$},
{\it JINST} {\bf 8} (2013) C12042.

\bibitem{Aune:2014sa}
S.~Aune {\it et al.},
\emph{Low background X-ray detection with Micromegas for axion search},
{\it JINST} {\bf 9} (2014) P01001.

\bibitem{admx2} 
S.~J.~Asztalos {\it et al.} 
\emph{An improved RF Cavity Search for Halo Axions},
\emph{Phys. Rev. D} {\bf 69} (2004) 011101 [astro-ph/0310042].

\bibitem{Raffelt:2008ggr}
G.G.~Raffelt,
\emph{Astrophysical axion bounds}
\emph{Lect. Notes Phys.} {\bf 741} (2008) 51 [hep-ph/0611350].

\bibitem{Shilon:2012} 
I.~Shilon, A.~Dudarev, H.~Silva and H.~H.~J.~ten~Kate,
\emph{Conceptual Design of a New Large Superconducting Toroid for IAXO, the New International AXion Observatory},
\emph{IEEE\ T \ Appl.\ Supercon.\ } {\bf 23}, 3, p. 4500604 (2013). 

\bibitem{Madsen:2009kkm}
K.K.~Madsen {\it et al.},
\emph{Optimizations of Pt/SiC and W/Si multilayers for the nuclear spectroscopic telescope array},
\emph{SPIE} {\bf 7437} (2009) 743716-1.

\bibitem{Meyer:2013pny}
  M.~Meyer, D.~Horns and M.~Raue,
  \emph{First lower limits on the photon-axion-like particle coupling from very high energy gamma-ray observation},
  \emph{Phys.\  Rev.\  D} {\bf 87}, 035027 (2013) [arXiv:1302.1208].

\bibitem{Isern:2010}
  J.~Isern, E.~Garcia-Berro, L.~G.~Althaus and A.~H.~Corsico,
  \emph{Axions and the pulsation periods of variable white dwarfs revisited},  
  \emph{Astron. Astrophys.} {\bf 512}, A86 (2010) [arXiv:1001.5248].
  
\bibitem{alpsII}
  R.~B\"ahre {\it et al.},
  \emph{Any light particle search II - Technical Design Report},
  \emph{JINST} {\bf 8} (2013) T09001,
  [arXiv:1302.5647 [physics.ins-det]]. 

\bibitem{hess}
P.~Brun, D.~Wouters (for the H.E.S.S. collaboration),
\emph{H.E.S.S. contributions to the 33rd International Cosmic Ray Conference (ICRC2013)},
\emph{proceedings of the 33rd ICRC, 2013, Rio de Janeiro}.
\end{thebibliography}
\end{document}